\documentclass[runningheads]{llncs}
\usepackage[utf8]{inputenc}
\usepackage[T1]{fontenc}
\usepackage[english]{babel}

\usepackage{graphicx}
\usepackage{mathtools}
\usepackage{array}
\usepackage{listings}
\usepackage{enumerate}
\usepackage{caption}
\usepackage{amssymb}
\usepackage{setspace}
\usepackage{tcolorbox}
\usepackage{authblk}
\usepackage{subfigure}
\usepackage{color}

\input{Qcircuit}
\usepackage{algorithm}
\usepackage[noend]{algpseudocode}

\begin{document}
\title{Function Maximization with \\ Dynamic Quantum Search}
\author{Charles Moussa\inst{1,2}\orcidID{0000-0002-5387-564X} \and
Henri Calandra\inst{3}, \\ 
and Travis S.~Humble\thanks{This manuscript has been authored by UT-Battelle, LLC, under Contract No. DE-AC0500OR22725 with the U.S. Department of Energy. The United States Government retains and the publisher, by accepting the article for publication, acknowledges that the United States Government retains a non-exclusive, paid-up, irrevocable, world-wide license to publish or reproduce the published form of this manuscript, or allow others to do so, for the United States Government purposes. The Department of Energy will provide public access to these results of federally sponsored research in accordance with the DOE Public Access Plan.}\inst{2}\orcidID{0000-0002-9449-0498}}
\authorrunning{C. Moussa et al.}
%
\institute{TOTAL American Services Inc.,
Houston, Texas USA
\and
Oak  Ridge  National  Laboratory,
Oak  Ridge,  Tennessee USA
\and 
TOTAL SA,
Courbevoie, France
}
\maketitle              
\begin{abstract}
Finding the maximum value of a function in a dynamic model plays an important role in many application settings, including discrete optimization in the presence of hard constraints. We present an iterative quantum algorithm for finding the maximum value of a function in which prior search results update the acceptable response. Our approach is based on quantum search and utilizes a dynamic oracle function to mark items in a specified input set. As a realization of function optimization, we verify the correctness of the algorithm using numerical simulations of quantum circuits for the Knapsack problem. Our simulations make use of an explicit oracle function based on arithmetic operations and a comparator subroutine, and we verify these implementations using numerical simulations up to 30 qubits. 

\end{abstract}
\section{Introduction}
Finding the maximal value in a poorly characterized function is a challenging problem that arises in many contexts. For example, in numerical simulations of particle dynamics, it is often necessary to identify the strongest interactions across many different particle trajectories. Prior results in quantum computing for unstructured search have shown that a quadratic speed up is possible when searching a poorly characterized function, and in this work, we consider an application of quantum search to the case of finding a maximal function value. 
\par
Quantum search was proposed originally to identify a marked item by querying an unstructured database \cite{quant-ph/9605043}. By using an oracular implementation of the database function, Grover proved that a quadratic speedup in unstructured search could be obtained relative to brute force search by using superposition states during the querying phase. These ideas have been applied to a wide variety of application contexts including function maximization and minimization. In particular, Durr and Hoyer have shown how to design an iterative version of quantum search to find the minimizing argument for an unspecified oracle \cite{durr1996quantum}. Their approach uses an updated evaluation of the oracle based on prior measurement observations, namely, to identify the smallest observed value. More recently, Chakrabarty et al.~have examined the problem of using a dynamic Grover search oracle for applications in recommendation systems \cite{Chakrabarty2017}. In addition, Udrescu et al.~have cast the application of quantum genetic algorithms into  a variant of dynamic quantum search \cite{Udrescu2006ImplementingQG}. 
\par
In this contribution, we implement the principles of dynamic quantum search for the case of function maximization. We use the Knapsack 0/1 problem as an illustrative example, which is a constraint problem that finds the largest weighted configuration for a set of possible items.  We extend recent work from Udrescu et al.~\cite{Udrescu2006ImplementingQG} by using quantum circuits for arithmetic operations. Our implementation uses arithmetic operations to build the dynamic oracle for quantum search function maximization \cite{GroverMax}. We test this implementation with a quantum program simulated using the Atos Quantum Learning Machine, which enables verification of the accuracy of the algorithm and estimation of the resources needed for its realistic implementation.  
\par
The paper is organized as follows: Sec.~\ref{sec:alg} formulates the quantum algorithm for function maximization using an iterative variant of quantum search, while  Sec.~\ref{sec:sim} presents the explicit implementation of these ideas using quantum arithmetic circuits and results from numerical simulation of those circuits. Finally, we explain how quantum search is used in this context and show how we simulated the example step by step in Sec.~\ref{sec:fin}.
\section{Quantum Maximization of Dynamic Oracle}
\label{sec:alg}
We show how to realize quantum search for function maximization using a dynamic oracle.  As an illustrative example, we consider a Knapsack 0/1 problem introduced in Udrescu et al.~as quantum version of a binary genetic algorithm called the reduced quantum genetic algorithm (RQGA) \cite{Udrescu2006ImplementingQG}. In particular, the genetic algorithm underlying RQGA was reduced to Grover search, thus removing the need for crossover and mutation operations for creating and selecting new candidate solutions. Rather, the complete set of candidate solutions is realized as an initial superposition over all the possible binary strings, e.g., using a Hadamard transform to prepare a uniform superposition state. We extend these ideas by developing an implementation using dynamic quantum search to solve the Knapsack problem.
\par
Our method begins with an initial uniform superposition state of the candidate solutions stored in an $n$-qubit register $q$, where the size is determined by the number of possible items. The initial superposition is evaluated with respect to an objective function noted as $f$. For our example, the function $f$ computes the fitness of the candidate solution. The computed value for the fitness is stored in a second quantum register denoted as $f$. The register of size $p$ stores this value using two's complement form, where the size is determined by the largest value of all items in the database. The fitness is computed explicitly using a unitary operator $U_f$ what applies to both quantum registers $q$ and $f$, i.e.,
\begin{equation}
(1/\sqrt{2^n}) \sum_{i=0}^{2^n-1} \ket{i}_{q} \ket{0}_{f} \rightarrow (1/\sqrt{2^n}) \sum_{i=0}^{2^n-1} \ket{i}_{q} \ket{f(i)}_{f}
\end{equation}
\par
After preparing a uniform superposition of the candidate solutions and associated fitness values, the amplitudes with a fitness value greater than a current maximum value are marked by an oracle operator. The oracle operator is implemented explicitly below using a comparator circuit that takes as input the computed fitness register and a classical threshold value. The amplitude of all marked items are subsequently amplified to increase the probability to measure a better candidate. In practice, we extend the definition of the fitness operator to exclude certain candidates in the superposition as being valid or invalid candidate. For example, when valid candidates must have a positive fitness value, then a single register element may be used to indicate mark values that are negative and therefore invalid. Theoretically, Grover's search requires $\approx13.6\sqrt{M}$ steps to achieve an error rate of less than 0.5 with $M$ being the size of the list searched \cite{GroverMax}.
\begin{figure}
\centering
\[
\Qcircuit @C=1.0em @R=.7em {
\lstick{q} & \gate{\mathcal{S}} & \ctrl{1} & \ctrl{2} & \qw & \qw & \qw & \qw & \ctrl{2} & \ctrl{1} & \gate{\mathcal{G}} & \qw\\
\lstick{w} & \qw & \multigate{1}{\mathcal{W}} & \qw & \multigate{3}{\mathcal{V}} & \qw  & \qw & \qw & \qw & \multigate{1}{\mathcal{W^{\dag}}} & \qw & \qw \\
\lstick{g} & \qw & \ghost{\mathcal{W}}& \multigate{1}{\mathcal{F}} & \ghost{\mathcal{V}} & \multigate{1}{\mathcal{I}} & \multigate{3}{\mathcal{C}}  & \multigate{1}{\mathcal{I^{\dag}}} & \multigate{1}{\mathcal{F^{\dag}}}  & \ghost{\mathcal{W^{\dag}}} & \qw & \qw  \\
\lstick{f} & \qw & \qw & \ghost{\mathcal{F}}  & \qw & \ghost{\mathcal{I}} & \ghost{\mathcal{C}} & \ghost{\mathcal{I^{\dag}}}  & \ghost{\mathcal{F^{\dag}}} & \qw & \qw & \qw\\
\lstick{v} & \qw & \qw & \qw & \ghost{\mathcal{V}} & \ctrl{-1}  & \qw  & \ctrl{-1} & \qw & \qw & \qw & \qw\\
\lstick{r} & \qw & \qw & \qw & \qw & \qw & \ghost{\mathcal{C}} & \qw & \qw & \qw & \qw & \qw\\
}
\]

    \caption{This quantum circuit diagram represents the algorithm for function maximization using iterative Grover search with a dynamic oracle. 
    	We only represent with one Grover iteration for simplicity (more would require repeating the steps after creating the superposition). 
    	The quantum registers $q$, $w$, $g$, $v$, and $f$ define respectively storage space for the candidate solutions, computed weight, garbage, validity, and fitness values. Gates in the circuit are $\mathcal{S}$ for preparing the initial superposition of candidate states, $\mathcal{W}$ for computing candidate weight, $\mathcal{F}$ for computing candidate fitness value,$\mathcal{V}$ for computing validity (by loading the maximum weight in register $g$ and using a comparator with register $g$) , $\mathcal{I}$ for inversing fitnesses of invalid candidates, $\mathcal{C}$ for comparing to the current maximum (after being loaded in register $g$), $\mathcal{G}$ represents the inversion about average operator. Measurement of the $q$ register should return a candidate that has a higher fitness than the current maximum loaded in $g$. }
    \label{fig:circuit}
\end{figure}
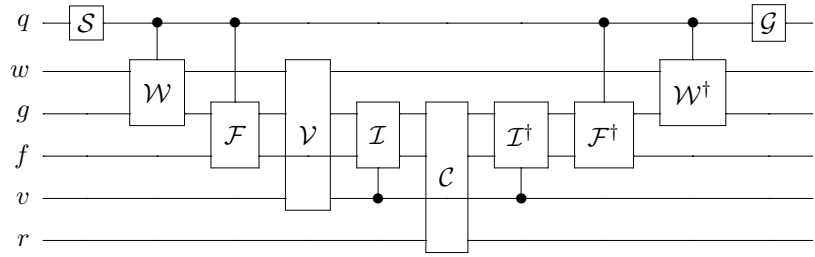
\par
Given this general overview of the algorithm, we next describe how to apply these ideas in a specific example for the Knapsack problem. Consider a backpack of $n$ possible items, where each item has a value and a weight. We seek the set of items that maximize the total value of the backpack while also not exceeding a defined total weight. Udrescu et al.~\cite{Udrescu2006ImplementingQG} have used the example of $n=4$ items, with the following properties: Item 1 (7 kg, \$40), Item 2 (4 kg, \$100), Item 3 (2 kg, \$50), and Item 4 (3 kg, \$30). The maximum weight allowed is 10 kg. Table~\ref{tab:items} summarizes the set of all possible solutions for this example, from which it is apparent that the optimal solution is the set of all items excluding Item 1. 
\begin{table}[ht]
	\centering
	\caption{Candidate solutions and values for Knapsack problem data.}
	\label{tab:items}
	\begin{tabular}{|c|r|r|r|r|}
		\hline
		\bf Candidate & \bf Fitness & \bf Weight &  \bf Validity  \\ \hline
		
		0 0 0 0 & 0 & 0 & valid  \\ \hline
		0 0 0 1 & 30 & 3 & valid  \\ \hline
		0 0 1 0 & 50 & 2 & valid  \\ \hline
		0 0 1 1 & 80 & 5 & valid  \\ \hline
		
		0 1 0 0 & 100 & 4 & valid  \\ \hline
		0 1 0 1 & 130 & 7 & valid  \\ \hline
		0 1 1 0 & 150 & 6 & valid  \\ \hline
		\textbf{0 1 1 1} & \textbf{180} & \textbf{9} & \textbf{valid}  \\ \hline
		
		1 0 0 0 & 40 & 7 & valid  \\ \hline
		1 0 0 1 & 70 & 10 & valid  \\ \hline
		1 0 1 0 & 90 & 9 & valid  \\ \hline
		1 0 1 1 & 120 & 12  & invalid  \\ \hline
		
		1 1 0 0 & 140 & 11 & invalid  \\ \hline
		1 1 0 1 & 170 & 14 & invalid  \\ \hline
		1 1 1 0 & 190 & 13 & invalid  \\ \hline
		1 1 1 1 & 220 & 16 & invalid  \\ \hline
	\end{tabular}
\end{table}
\par
We summarize the computational steps taken to solve this example using the dynamic quantum search method within the quantum circuit model. Several quantum registers are used to store and process the problem input data, and we consider all registers to be initialized in the $\ket{0}$ state.
\begin{itemize}
	\item $q$: an $n$-qubit register for storing candidate solutions. $n=4$ in this example.
	\item $f$: a $p$-qubit register for storing the value of the fitness calculation in two's complement representation, where $p$ is sufficient to store the sum of all possible values. $p = 6$ is this example.
	\item $w$: a register for storing the total mass of a candidate backpack. $\dim(w) = 5$  using unsigned integer representation in this example.
	\item $g$: a register for storing intermediate computational states and numerical constants. $\dim(g) = 6$ for this example.
	\item $v$: a 1-qubit register to mark a candidate fitness as valid or invalid.
	\item $r$: a 1-qubit register used by the oracle.
\end{itemize}
\par
Following the declaration and initialization of all registers, the candidate register $q$ is transformed under the $n$-fold Hadamard operator to prepare an uniform superposition of possible solution states. An initial fitness value threshold is selected by random number generation. We then apply a composite operator denoted as $o(x)$ that consists of the following stages: 
\begin{enumerate}
	\item Calculate the total mass, fitness, and validity of the candidates in registers $w$, $f$, and $v$.
	\item Compare the fitness register $f$ against the current threshold value stored in $g$. If the fitness value is greater than the current threshold value, then set $r$ with the effect of marking the states, that is having the effect $ \ket{-} \rightarrow (-1)^{o(x)} \ket{-} $ on the qubit oracle.
	\item Uncompute the register $f$, $w$, and $v$.
	\item We finally measure (after application of Grover iterations) the candidate solution register $q$ (and fitness register $f$ if we apply again the fitness computation circuit).
\end{enumerate}

\par
When computing the total mass and fitness, we store the weight and value of each item in the register $g$. Using a controlled adder, we perform addition with this register to store the sum of all weights and values for each candidate solution. As an illustration, consider the case that the fitness and mass are initially zero. We begin by adding the mass and value of the first item to the register. In our example problem, Item 1 has a mass 7 kg and value 4\$. We load the unsigned integer mass 7 (000111) into the workspace register $g$. The first qubit of the register $q$ acts as a control register for adding the mass of Item 1 in the register $m$ using controlled addition circuit. Similarly, we add the value 4 to the fitness register $f$ for those candidate solutions containing Item 1. This sequence of steps prepare the following quantum state (omitting workspace registers for clarity),
\begin{equation}
\sum_{i=0}^{7} \ket{i}_q \ket{00000}_w \ket{000000}_f + \sum_{i=8}^{15} \ket{i}_q \ket{00111}_w \ket{000100}_f 
\end{equation}
The remaining items in the database are treated similarly.
\par
As shown in Table~\ref{tab:items}, some candidate solutions have a total weight exceeding the constraint condition. These candidates are flagged by setting the register $v$ to 1, a step implemented by comparing the maximum weight constraint with the value of the register $w$. If the candidate weight exceeds the constraint, the register $v$ is flipped and we inverse their values to put them into their negative form so that when comparing to the current maximum, they will not be marked using the comparator quantum circuit operator.
The negation operation is controlled by the validity qubit and implemented by complementing the fitness register $f$ and adding 1 to it as it is done in two's complement arithmetic.
\begin{equation}
\left(\sum_{i \in valid} \ket{i}_q \ket{m(i)}_w \ket{f(i)}_f \ket{0} + \sum_{i \in invalid} \ket{i}_q \ket{m(i)}_w \ket{-f(i)}_f \ket{1}_v \right) \ket{001010}_g
\end{equation}
\par
At this stage, we wish to mark those candidate states that are valid and uncompute all values other than the candidate register. After preparing a superposition state representing valid possibilities, we use amplitude amplification to other registers as $\ket{0}$ and the qubit oracle as $ \ket{-} $: 
\begin{equation}
\sum_x \ket{i}_q (-1)^{o(i)} \ket{-}_r
\end{equation}

We provide a pseudo-code representation for the steps before amplitude amplification in the Appendix.
Grover iterations are finally used to change amplitudes. The number of required Grover iterations depends on the number size of the database as well as the number of actual solutions $M$ as
\begin{equation}
\left\lceil \frac{\pi}{4} \sqrt{\frac{N}{M}} \right\rceil
\end{equation}
In our search for the maximizing solution, we do not know the number of solutions at each iteration in a maximization process with this quantum algorithm for finding the maximum. We therefore must use a version of quantum search that circumvents the need to estimate $M$. Boyer et al.~\cite{quant-ph/9605034} have provided methods for ensuring probabilistic success of the algorithm in the absence of knowledge about the number of items.
\par
\textbf{Quantum search for the case $M$ is unknown}
\begin{enumerate}
	\item Initialize $m=1$ and set a constant $ \lambda = 6/5 $.
	\item Choose a random integer $j$ that is less than $m$.
	\item Apply $j$ Grover routines over the superposition of candidates
	\item In our case, apply again the part 1 of the oracle to have a fitness and the validity.
	We can also do this part with a classical function.
	\item Measure to get an outcome candidate/fitness/validity.
	\item If we get a valid solution and the fitness is greater than the current max, we stop and return them as new solution and
	new current max.
	\item Else set $m$ to min($\lambda m, \sqrt{N}$) and repeat from step 2.
\end{enumerate}
We may not have a solution in the case we already have the maximum of the problem but an appropriate time-out or early stopping can be set and the complexity would remain $\sqrt{N}$. 
\par 
As an illustration, consider the case that the current threshold is 13. Candidate states in the superposition for our example problem will be marked if their associated weight is strictly greater than 13, as is the case for candidates 0110 and 0111. The resulting superposition becomes
\begin{equation}
\frac{1}{4} \left(\sum_{i \ne 0110,i \ne 0111} \ket{i}  - \ket{0110} - \ket{0111} \right)
\end{equation}
The result of the Grover search prepares the state
\begin{equation}
\frac{1}{8} \left(\sum_{i \ne 0110,i \ne 0111} \ket{i}\right)  - \frac{5}{8} \left(\ket{0110} + \ket{0111} \right)
\end{equation}
which yields acceptable candidate solutions with a probability $|5/8|^2 \approx 0.39$. 
\section{Quantum Circuit Simulation}
\label{sec:sim}
We describe our implementation of these algorithm using integer arithmetic with unsigned integers presented in a binary representation and signed integers presented as two's complement. Let $a = (a_0,a_1,...,a_{n-1})$ be the $n$-bit representation of the integer $A$, such that $A = \sum_i a_i 2^i \in [0, 2^n - 1]$. For signed integers, one bit is used for the sign and we can represent numbers $[-2^{n-1}, 2^{n-1}-1]$. The most significant bit represents the sign, which is `$+$' for 0 and `$-$' for 1. Given two integers $a$ and $b$, bitwise addition is used to calculate the resultant $a+b$. For subtraction, recall that $a-b = (a'+b)'$ where $x'$ is the bit-wise complement of $x$ obtained by flipping all bits. Usually, an extra bit is required for addition but when not used, we actually do modular addition. To compare two numbers $a$ and $b$, one can do the subtraction and look at the last bit of the subtraction result. When it is 1, $a<b$.
\par
\begin{figure}[!h]
	\centering
	\includegraphics[width=0.6\columnwidth]{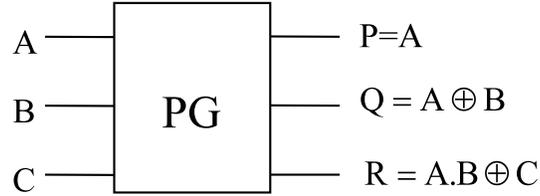}
	\caption{Peres gate representation.}
	\label{fig:peres}
\end{figure}

The quantum ripple adder acts on inputs $a$ and $b$ by encoding those binary representations as the $ \ket{a,b} $. The resulting output from in-place addition is the state $ \ket{a,a+b} $. Note that we use another bit called high bit as the addition result may exceed $ 2^n $. In Ref.~\cite{Thapliyal:2013:DER:2533711.2491682} a quantum circuit without ancilla qubit and with less quantum operations than previously proposed circuits was designed. This approach uses the Peres gates described in Fig.~\ref{fig:peres}.
\par
Cuccaro et al.~previously described how the quantum ripple-adder may be adapted for other arithmetic tasks \cite{RippleAdder}, and we make use of these implementations for our oracle definitions. For example, we define a subtraction method using the complement technique mentioned above.
For modular addition, we consider the last significant bit of the number $b$ as the high bit, and apply the adder on the $n-1$ first bits of $a$ and $b$, and apply a CNOT onto qubits representing the last significant bits. For a comparator, we use only the part of the adder where we compute the high bit and uncompute. If the high bit is 1, $a < b$. For a controlled adder, where given a control qubit we apply the adder if the control is $ \ket{1} $, a quantum circuit was proposed in \cite{MuozCoreas2017TcountOD}.
\par
For our simulations, we use the Atos Quantum Learning Machine, a numerical simulation environment designed for verifying quantum software programs. It is dedicated to the development of quantum software, training and experimentation. We use the QLM as a programming platform and a high-performance 
quantum simulator, which features its own universal quantum assembly programming language (AQASM, Atos Quantum Assembly Language) and a high-level quantum hybrid language ({PyQASM}) embedded in the popular Python language. The QLM is capable of performing exact simulations of up to about 30 qubits using a  vector-based representation of register states. The simulator manages the execution of the programmed sequence of quantum gates.
\par
For our small example instance, we can find the optimal backpack in two Grover iterations (or two circuits of one Grover iteration).
Some instances require more due to the probabilistic measurement outcomes, but a typical execution converges to valid solutions. It is possible to use the suboptimal solution as the threshold for the next instance of the algorithm in order to verify if better solutions are available. We just need a set of $n$ qubits to represent all possibilities while classically, this would be represented by many bit strings to form a population of candidates that will change over iterations like it is done in genetic algorithms.
\section{Conclusion}
\label{sec:fin}
In this contribution, we have shown how to implement function maximization using dynamic quantum search. Our approach has applied an explicit oracle for the quantum search algorithm and implemented a quantum algorithm for binary optimization of the Knapsack problem. Our implementation illustrates the idea of evaluating the fitness of all possible candidate solutions. However, our approach does require uncomputation of any ancillary register to ensure the correctness of the Grover iteration. While the added complexity is manageable, it would be advantageous to develop alternative search methods that act on subspaces of entangled registers. We anticipate that such methods would avoid the need for uncomputation of the ancillary registers and, in our example, permit more efficient steps to iteratively isolate the maximal fitness value.
\section*{Acknowledgments}
CM and HC acknowledge support from Total and TSH acknowledges support from the U.S.~Department of Energy, Office of Science, Early Career Research Program. Access to the Atos Quantum Learning Machine was provided by the Quantum Computing Institute at Oak Ridge National Laboratory.

\appendix

\section{Pseudo-code for the Knapsack example}

\begin{algorithm}
	\caption{Pseudo-code of the quantum circuit for the steps before amplitude amplification.}\label{euclid}
	\begin{algorithmic}[1]
		
		\Require{Current maximum value, quantum circuit with all qubits prepared in $\ket{0}$.\newline }

		\State Apply Hadamard Transform on register $q$ to get all possible candidates in superposition.
		\State Set register $r$ as $\ket{-}$.\newline 
		
		\Comment{Compute the weight $w_i$ of the candidates.}
		\State Load $w_1$ in register $g$.
		\State Control-add $w_1$ into register $w$ if the first qubit of register $q$ is 1.
		\State Load $w_2$ in register $g$.
		\State Control-add $w_2$ into register $w$ if the second qubit of register $q$ is 1.
		\State Load $w_3$ in register $g$.
		\State Control-add $w_3$ into register $w$ if the third qubit of register $q$ is 1.
		\State Load $w_4$ in register $g$.
		\State Control-add $w_4$ into register $w$ if the fourth qubit of register $q$ is 1.\newline 
		
		\Comment{Compute the fitness $f_i$ of the candidates.}
		\State Load $f_1$ in register $g$.
		\State Control-add $f_1$ into register $f$ if the first qubit of register $q$ is 1.
		\State Load $f_2$ in register $g$.
		\State Control-add $f_2$ into register $f$ if the second qubit of register $q$ is 1.
		\State Load $f_3$ in register $g$.
		\State Control-add $f_3$ into register $f$ if the third qubit of register $q$ is 1.
		\State Load $f_4$ in register $g$.
		\State Control-add $f_4$ into register $f$ if the fourth qubit of register $q$ is 1.\newline 
		
		\Comment{Define validity of the candidates.}
		\State Load the maximum weight (10 kg in the example) in register $g$. 
		\State Use a comparator with register $w$ to define the validity of a candidate in $v$.
		\State Inverse the fitness $f$ of the invalid candidates (for which $v$ is 1).\newline 
		
		\Comment{Mark better candidates.}
		\State Load current maximum in register $g$.
		\State Use a comparator with register $f$ to mark better candidates than the current one in register $r$.
		
		\State Uncompute.

	\end{algorithmic}
\end{algorithm}

%
%
%
\bibliographystyle{splncs04}
\bibliography{articlerefs}

\end{document}